\documentclass{aa}
\input psfig.tex
\usepackage{graphicx}
\usepackage{natbib}

\usepackage{array}
\usepackage{graphics}
\usepackage{latexsym}
\usepackage{amssymb}
\usepackage{amsmath}
\usepackage{fancyhdr}
\usepackage{morefloats}

\bibpunct{(}{)}{;}{a}{}{,}
\include{hyphe}
\begin{document}

\newcommand{\ROBO}[0]{RO\-BO}
\newcommand{\NTSPH}[0]{NB-\-TSPH}
\newcommand{\evol}[0]{\textsc{EvoL}}
\newcommand{\nel}[1]{n_\mathrm{#1}}
\newcommand{\ith}[1]{\emph{#1}-th }
\newcommand{\commento}[1]{}

\title{MaNN: Multiple Artificial Neural Networks \\ for modelling
the Interstellar Medium}

\author{Tommaso Grassi$^{1}$, Emiliano Merlin$^{1}$, Lorenzo Piovan$^{1}$, Umberto Buonomo$^{1}$, Cesare Chiosi$^{1}$}

\institute{$^1$ Department of Astronomy, Padova University,
  Vicolo dell'Osservatorio 3, I-35122, Padova, Italy\\
  \email{{tommaso.grassi\char64unipd.it, cesare.chiosi\char64unipd.it}}}
\date{Received: March 2011; Revised: *** ***; Accepted: *** ***}

\abstract {} {{Modelling} the complex physics of the
Interstellar Medium (ISM) in the context of large-scale numerical
simulations is a challenging task. A number of methods have been
proposed to embed a description of the ISM into different codes. We
propose a new way to achieve this task: Artificial Neural Networks
(ANNs).} {The ANN has been trained on a pre-compiled model database,
and its predictions have been compared to the expected theoretical
ones, finding good agreement both in static and in dynamical tests
run using the Padova Tree-SPH code \textsc{EvoL}.} {A neural network
can reproduce the details of the interstellar gas evolution,
requiring limited computational resources. We suggest that such an
algorithm can replace a real-time calculation of mass elements
chemical evolution in hydrodynamical codes.} {} 

\titlerunning{Multiple Artificial Neural Networks for modelling
the ISM}
\authorrunning{Grassi, Merlin, Piovan, Buonomo \& Chiosi }
\maketitle

\section{Introduction}\label{Introduction}

In the framework of modern large-scale numerical hydrodynamical
simulations, the correct modelling of the Interstellar Medium (ISM)
plays an increasingly crucial role as many physical phenomena such
as star formation, energy feedback, diffusion of heavy elements and
others, strongly depend on {it and} simultaneously affect the
properties of the ISM. The task is, however, very demanding. This is
ultimately due to the complexity of the hydrodynamical interactions
of {the} Baryonic Matter in large scale systems. Contrary to
the simple behaviour of {the} Dark Matter dynamically
governed only by gravity, the behavior of {the} Baryonic
Matter is the result of many complex and interwoven physical
phenomena. Therefore, to obtain a realistic model of the ISM,
{many physical} processes must be taken into account going
from  chemical reactions among atoms and molecules occurring in
complicated networks, dust formation and destruction, chemical
reactions taking place on dust grains, to heating and cooling by
various mechanisms, photo-ionization and radiative transfer. These
processes span wide ranges of temperature and {density}, thus
making their accurate, simultaneous description a cumbersome affair.

Moreover, even when the physics is described with the desired
accuracy, the number of mass elements (particles) customarily used
in large-scale simulations is large (typically $10^4$ to $10^7$), so
that large amounts of computational workload are requested. Clearly,
computing the internal evolution of each mass element in detail is
extremely time-consuming, whatever the method chosen to follow it
might be.

The aim of this study is to present a method able to alleviate part
of the above difficulties by approximating the exact solution of a
detailed physical model and yet ensuring both a robust and
sufficiently precise description of the gas properties, while
maintaining the computational workload to an acceptable level.

The method stands on the model of the ISM and companion code
\textsc{Robo} developed by \citet{Grassi10}. In brief,
\textsc{Robo} follows the evolution of a gas element during a given
time interval $\Delta t$. In other words, given the initial physical
and chemical state of a  gas element as functions of {its
hydrodynamical properties, the relative abundances} of its chemical
constituents and the kind of interactions with the surrounding
medium, the model {calculates} the \textit{final} state after
some time interval $\Delta t$.

The straight implementation of \textsc{Robo} into a generic
hydrodynamical simulation (and companion code) to perform the
on-the-fly computation of the ISM properties, although desirable,
would be far too expensive in terms of computational costs. The
easiest way of proceeding would be to store the results of
\textsc{Robo} as look-up tables (grids) where the values of interest
(e.g. the physical parameters of the gas element: mass fractions of
molecular, neutral, ionized hydrogen; temperature, density,
pressure; the abundance of {dust}, and so on) are memorized
for a given (large) set of initial configurations. However, each
parameter to be stored corresponds to one dimension of the look-up
table (grids). Thus, tracking $N$ parameters requires a
$N$-dimensional matrix. Such matrix is discrete, so an interpolation
in a $N$-dimensional space is required to know the final state
corresponding to a given initial state of a generic gas element. The
all procedure when applied many times to a high number of gas
particle would be very time consuming. A different strategy must be
conceived.

To cope with this, we develop an algorithm based on the Artificial Neural Networks (ANN)
technique to establish the correspondence between the
\textit{initial} $+$ \textit{final} state. The ANN technique can
handle a large number of parameters without requiring a real-time
interpolation. The {ANNs require} few floating point
operations and {``learn''} to behave like the original
physical models on which they are trained. The computational cost of
it is easily affordable.

The paper is organized as follows. In Section \ref{robo} we briefly
describe the ISM model and companion code \textsc{Robo}. In Section
\ref{ann} we briefly review the concept and algorithm of {the
ANN} in general, whereas {in Section} \ref{MaNN} we present
the ANN we have set up for our purposes. This is named
\textmd{MaNN}. In the
same section, we also examine the ability of \textmd{MaNN} in
reproducing  the data calculated by \textsc{Robo}. In Section
\ref{testing} we describe how \textmd{MaNN} is implemented  into the
NB-TSPH code \textsc{EvoL} of \citet{Merlin2010} and present a
{suitable} hydrodynamical test aimed at checking the
robustness of the method. Finally, in Section \ref{conclusions} we
summarize our results and draw some general considerations.

\section{\textsc{Robo}: the ISM model and companion code}\label{robo}

\textsc{Robo} is a numerical code specifically designed  to study
the evolution of the ISM. It includes several atomic and molecular
species linked together by a large network of reactions. In the
following we summarize the main features of \textsc{Robo} and
emphasize the fact that  \textsc{Robo} is specifically tailored to
follow the evolution {of the} coolants owing to their
important role in the wider context of the ISM evolution. To this
purpose, we follow the temporal variation of molecules like $\mathrm
H_2$, $\mathrm{HD}$ and metals like C, O, Si, Fe and their ions. For
all the details the reader should refer to \citet{Grassi10}.

\textsc{Layout of the ISM model}. The model deals with an ideal ISM
element of unit volume, containing gas and dust in arbitrary initial
proportions, whose initial physical conditions are specified by a
set of parameters, which is let evolve for a given time interval.
The history leading the element to that particular initial physical
state is not of interest here. The ISM element is mechanically
isolated from the host environment, i.e. it does not expand or
contract under the action of large scale forces, however it can be
interested by the passage of shock waves originated by physical
phenomena taking place elsewhere (e.g. supernova explosions).
Furthermore, it  does not acquire nor lose material, {so
that} the conservation of {the} total mass applies, even if
its chemical composition can change with time. It is immersed in a
bath of UV radiation generated either by nearby or internal stellar
sources and in a field of cosmic ray radiation. It can generate its
own radiation field by internal processes and so it has its own
temperature, density and pressure, each other related by an Equation
of State (EoS). If observed from outside, it would radiate with a
certain spectral energy distribution. For the aims of this study, we
do not need to know the whole spectral energy distribution of the
radiation field pervading the element, but only the UV component of
it. Given these hypotheses and the initial conditions, the ISM
element {evolves} toward another physical state under the
action of the internal network of chemical reactions changing the
relative abundances of the elemental species and molecules, the
internal heating and cooling processes, the UV radiation field, the
field of cosmic rays, and the passage of shock waves. The model is like an operator
determining in the space of the physical parameters conditions the
vector field of the local transformations  of the ISM from one
initial state to a final one. This is the greatest merit of this
approach, which secures the wide applicability of the model.

{The chemical network}. The following species are tracked by
\textsc{Robo}: $\mathrm{H}$, $\mathrm H^+$, $\mathrm H^-$,
$\mathrm H_2$,
$\mathrm{D}$, $\mathrm D^+$,  $\mathrm D^-$,
$\mathrm{D_2}$, $\mathrm{HD}$, $\mathrm{HD}^+$,
 $\mathrm{He}$, $\mathrm{He}^+$, $\mathrm{He}^{++}$,
$\mathrm{C}$, $\mathrm C^+$, $\mathrm{O}$, $\mathrm O^+$,
$\mathrm{Si}$, $\mathrm{Si}^+$,  $\mathrm{Fe}$, $\mathrm{Fe}^+$, and
$\mathrm e^-$. The list of reactions and their rates that are
included in \textsc{Robo} can be found in \citet{Grassi10}. The
reactions are of different type and can be grouped as follows
\begin{itemize}
  \item collisional ionization
    ($\mathrm A +\mathrm e^- \to \mathrm A^+ + 2\mathrm e^-$),
  \item photo-recombination
    ($\mathrm A^+ +\mathrm e^- \to \mathrm A + \gamma$),
  \item dissociative recombination
    ($\mathrm A^+_2 +\mathrm e^- \to 2\mathrm A$),
  \item charge transfer
    ($\mathrm A^+ +\mathrm B \to \mathrm A +\mathrm B^+$),
  \item radiative attachment
    ($\mathrm A +\mathrm e^- \to \mathrm A^- +\gamma$),
  \item dissociative attachment
    ($\mathrm A +\mathrm B^- \to \mathrm{AB} +\mathrm e^-$),
  \item collisional detachment
    ($\mathrm A^- +\mathrm e^- \to \mathrm A +2\mathrm e^-$),
  \item mutual neutralization
    ($\mathrm A^+ +\mathrm B^- \to \mathrm A +\mathrm B$),
  \item isotopic exchange
    ($\mathrm A^+_2 +\mathrm B \to \mathrm{AB}^+ +\mathrm A$)
  \item dissociations by cosmic rays
    ($\mathrm{AB} +\mathrm{CR}\to \mathrm A +\mathrm B$),
  \item neutral-neutral
    ($\mathrm{AB} +\mathrm{AB}\to \mathrm A_2+\mathrm B_2$),
  \item ion-neutral
    ($\mathrm{AB}^+ +\mathrm{AB}\to \mathrm{AB}_2^+ +\mathrm A$),
  \item collider
    ($\mathrm{AB} +\mathrm C \to \mathrm A +\mathrm B +\mathrm C$),
    \item
    ionizations by field photons ($\mathrm{A} +\gamma \to \mathrm{A^+} +\mathrm{e^-}$)
\end{itemize}
where $\mathrm A$ and $\mathrm B$ are two generic atoms, $\gamma$ is
a photon. Note that cosmic rays and photo-ionization effects are
included.

\textsc{Dust}. \textsc{Robo} also tracks the evolution of dust, in
turn made by several components, of which it follows the formation
and destruction. The formation of dust is modelled according to the
prescriptions by \citet{Dwek98}, however with some modifications
that allow us to know  when a species belonging to the gas phase is
captured  by dust (trapped at the surface of dust grain lattice).
The evolution of the dust type (both in abundance and dimensions) is
followed for {silicates and carbonaceous grains}. Dust is
also destroyed by thermal sputtering and shocks. The first process
is modeled using the description by \citet{DraineSalpeter79a}, that
depends on the environment density and the grain size. We also add
the dependence on temperature, considering that hot gas  disrupts
the dust grains more efficiently than the cold one. The dust
disruption by {shocks requires} the velocity distribution of
the gas component to be known. This is supposed to be in the
turbulence regime and the velocity field to be suitably described by
the Kolmogorov-law. This allows us to describe the shocks on very
small scales (contrary to what usually happens in large-scale
simulations that do not have the required spatial resolution). Dust
is an important component of the ISM because it affects the
formation of $\mathrm{H_2}$ (one of the most efficient coolants of
the ISM) and $\mathrm{HD}$  and also the gas phase catalyzing some
chemical processes. The formation of $\mathrm{H_2}$ and
$\mathrm{HD}$  is modeled as in \citet{CazauxSpaans09}, whose
description depends on the gas and dust temperature. Finally, the
photoelectric ejection of electrons from dust grains is described in
detail as well. We can estimate the amount of heating (and cooling)
produced by each size-bin of the grain distribution. For further
details, see \citet{Grassi10}.

\textsc{Cooling}. Several cooling processes are considered in
\textsc{Robo}. In the high temperatures regime ($T \ge 10^4$ K) the
metallicity dependent cooling rates of \citet{SutherlandDopita93}
are adopted. For temperatures lower than $10^4\,\mathrm{K}$ we
consider the following processes: (i) cooling by  molecular hydrogen
according to the formulation by  \citet{GalliPalla98}, however,
supplemented by the results of \citet{GloverJappsen2007} who take in
account the $\mathrm{H_2}-\mathrm{H}$ interaction and the collisions
with $\mathrm{He}$, $\mathrm{H^+}$, $\mathrm{H_2}$ and free
electrons; (ii) {cooling} by metals is modeled including C,
Si, O, Fe, and their ions as in \citet{Maio07},
\citet{GloverJappsen2007} and \citet{HollenbachMcKee89}; (iii)
{cooling} by deuterated molecular hydrogen according to the
model  by \citet{Lipovka05}; (iv) cooling by the
CO molecule and (v) finally, cooling by \citet{Cen92}.

The total cooling rate is the sum of all the contributions
\begin{displaymath}
\Lambda_{\mathrm{tot}}=\Lambda_\mathrm{SD}+\Lambda_{\mathrm H_2}+\Lambda_\mathrm{HD}+\Lambda_\mathrm{CO}
    +\Lambda_\mathrm{metals}+\Lambda_\mathrm{CEN}\,,
\end{displaymath}
where the various terms $\Lambda s$ are all functions of temperature
and density and  $\Lambda_\mathrm{SD}$ is the cooling rate by
\citet{SutherlandDopita93}, $\Lambda_{\mathrm H_2}$,
$\Lambda_\mathrm{HD}$ and $\Lambda_\mathrm{CO}$ are the cooling
rates  of the indicated molecular species, and
$\Lambda_\mathrm{metals}$ is the cooling by metals (C, O, Si and
Fe). Finally, $\Lambda_\mathrm{CEN}$ is the cooling by \citet{Cen92}.

It is worth to notice that in the tests presented in this paper the
cooling proposed by \citet{Cen92} is disabled (i.e. $\Lambda_\mathrm{CEN}=0$).

\textsc{Heating}. For the photo-dissociation of {the}
molecular hydrogen and {the} UV pumping, {the} H and
He photo-ionization, {the} $\mathrm{H_2}$ formation in the
gas and dust phase {and, finally, the} ionization from cosmic
rays, heating is described as in \citet{GloverJappsen2007}. For the
heating due to the photoelectric effect on dust grains, the model
proposed by \citet{BakesTielens94} {and \citet{WeinDraine01}}
is used.

\textsc{Integration time of the models}. Each model of the ISM is
followed during a total time of about $10^4\ \mathrm{yr}$. This
choice stems from the following considerations. Each model of the
ISM is meant to represent the thermal-chemical history of a unit
volume of ISM, whose initial conditions have been established at
certain arbitrary time and whose thermal-chemical evolution is
followed over a time scale comparable to the maximum time step of the
TreeSPH code.  
The initial physical conditions
are fixed by a given set of parameters each of which can vary over
wide ranges. One has to solve the network of equations for a time
scale long enough to reveal the variations due to important
phenomena such star formation, cooling and heating, but not long to
allow to the system to depart from the instantaneous situation one
is looking at. The value of $10^4$ yr resulted to be a good
compromise. The initial value of the time-step is $1\ \mathrm{yr}$.
This time-step determines the minimum number of steps required to
cover the time spanned by a model. It means that each simulation
needs at least $10^4$ iterations to be completed. The numerical
integrator may introduce shorter time steps depending on the
complexity of the problem. 
This value for the time-step seems to keep the
system stable during the numerical integration\footnote{In
\citet{Grassi10} the adopted total integration interval was about
$10^3$ longer, i.e. about $10^7$ yr, because {we were}
interested in exploring the chemical evolution of the ISM over a
sizable time scale, whereas now we are interested in the evolution
over a time scale comparable to that of typical time steps of
NB-TSPH simulations.}.

\textsc{Database of models}. Varying the initial value of the
parameters describing the ISM (such as the temperature, density,
abundances of elemental species etc.), \textsc{Robo} is used to
create a large number of evolutionary models.

\section{The Artificial Neural Networks}\label{ann}
As already mentioned, the ideal way of proceeding would be to insert
\textsc{Robo} into a code simulating  the temporal evolution of
large scale structures and to calculate the thermodynamical
properties of the ISM component. In all practical cases, this would
be unreasonably time consuming. To cope with it without loosing
accuracy in the physical description of the ISM we construct an ANN
able to replace \textsc{Robo} in all cases.

In brief, ANNs  are non-linear tools for statistical data modeling,
used to find complex relationships between input  and output data or
to discover  patterns in complex datasets. The first idea of an ANN
is by \citet{Hebb49}. In the early sixties \citet{Rosenblatt62}
built the first algorithm based on the iterative penalty and reward
method, and finally, after two decades of silence, new ideas were
brought by \citet{CarpenterGrossberg87a,CarpenterGrossberg87b,
CarpenterGrossberg90} (adaptive resonance theory),
\citet{Hopfield82, Hopfield84} (associative neural network) and
\citet{Rumelhart86, McClelland86}, who developed the
\textit{back-propagation algorithm}, today a common and widely used
training scheme.

Demanding computational problems (e.g. multidimensional fits or
classifications) can be easily solved using ANNs whereas other
methods would require large computing resources. An ANN is based on
a simple conceptual architecture inspired by the biological nervous
systems. {It looks like} a network composed by
\textit{neurons} and linked together by \textit{synapses} (numerical
weights). This structure is modeled with a simple algorithm,
designed to predict an \textit{output state} starting from an
original \textit{input state}. To achieve this, the ANN must be
\textit{trained}, so that it can ``learn'' to predict the output state
from  a given set of initial configurations. The {simplest
way} of doing it, is the \emph{supervised learning}:
input-output values are fed  to the network, then the ANN
tentatively computes an output set for each input set, and the
difference between the predicted and the original output (i.e. the
error) is used to calculate penalties and rewards to the synaptic
weights. If {the} convergence to a solution exists, after a
number of iterations, the predicted output value gets very close to
the original {one:} \textit{the ANN has learned}.

The network stores the original data in some dedicated neurons, the
so called \emph{input neurons}, elaborates these data and sends the
results to another group of neurons that are called \emph{output
neurons}. Thus, data flow from input to output neurons. There is
another group of neurons, named  \emph{hidden neurons}, that help
the ANN to elaborate the final output (see Fig. \ref{annfig}). Each
neuron (or unit) performs a simple task: it becomes \emph{active} if
its input signal is higher than a defined threshold. If one of these
units becomes active, it emits a signal to the other neurons; for
this reason, each unit can be considered as a \textit{filter} that
increases or decreases the intensity of the received signal. The
connections between neurons simulate the biological synapses and
this is the reason why they are called \emph{synaptic weights} or,
more simply, \emph{weights}.

The whole process can be written as follows:
\begin{equation}\label{activation}
    n_i=f(x) \left(\sum_jw_{ij}n_j-\theta_i\right)\,,
\end{equation}
where $n_i$ is the $i$-th neuron value, $f(x)$ is the activation
function,  $w_{ij}$ is the weight between $i$-th and $j$-th neuron,
$\theta_i$ is the neuron threshold, and $x$ is the signal. For the
activation function we adopt a \emph{sigmoid}
\begin{equation}
    f(x)=\frac{1}{1+e^{-\beta x}}\,,
\end{equation}
where $\beta$ is a parameter fixing the sigmoid slope. It is worth
recalling that other expressions for the activation function can be
used.

In general the ability of the network to reproduce the data
increases with the number of the hidden units up to a certain limit
otherwise the ANN gets too ``stiff'' so that predicting the original
output values gets more difficult. \emph{The ANN stops learning}.
Another reason to avoid many hidden neurons is that the procedure
becomes time consuming (one must remember that we are mainly dealing
with  with matrix products). On the other side, an ANN with too few
hidden units gets ``dull'' in reproducing the original data.
Therefore, there is an optimal number of hidden neurons for each
ANN.

There are several possible ANN architectures, depending on the task
the network is designed to perform. The already cited
back-propagation algorithm \citep{Rumelhart86} is among the most
versatile methods, {it is one of the most suited to the
supervised learning stage and can be applied to a wide range of
problems}. The name stems from the fact that the error produced by
the network is propagated back to its connections to change the
synapsis weights and consequently to reduce the error. As a
detailed description of the ANN technique is beyond the aims of this
paper, we limit ourselves to describe in some detail only the
particular algorithm we have adopted for our purposes.


\section{{\textmd{MaNN}}: aims and architecture}\label{MaNN}

\textmd{MaNN} is the network we built to reproduce the ISM
{output} models calculated by \textsc{Robo}. We use a layer
of input neurons where each input value corresponds to a free
parameter of the \textsc{Robo} models, i.e. the ones shortly
described in Sect. \ref{robo}. The definition of {the} model
in \textmd{MaNN} requires some {clarifications}. While
\textsc{Robo} calculates the whole temporal evolution of a model,
\textmd{MaNN} needs only the set of initial {parameters} (the
input model) and the set of final results ({the} output
model). Therefore, a \textmd{MaNN} model is made of two strings
(vectors) of quantities: the initial conditions for the key
parameters ({the} input vector) {and} the results for
the quantities describing the final {state, after that} the
assigned time interval has elapsed ({the} output vector). We
refer to the input vector as the vector $\bar{x}$ whose dimension,
i.e. {the} number of input units, is $n_x$. We then choose
the shape of the output layer of neurons having in mind an ``optimal''
number of parameters to be used in a typical dynamical simulation.
We use $\bar{y}$ to indicate these units and the number of the
neurons belonging to this layer is $n_y$ \footnote{ In practice, to
chose the input and output vectors and their dimensions we pay
attention to the characteristics of the code to which the
\textmd{MaNN} is applied, namely the Padova Tree-SPH code
\textsc{EvoL} by \citet{Merlin2010}. {The same reasoning can
be applied to} the code generating the input data, \textsc{Robo} in
our case. However, the whole procedure can be generalized to any
other type of codes.}. For example, if the code generating the input
data only tracks the abundance of neutral hydrogen and the
temperature, the \textmd{MaNN} input vector  will be
$\bar{x}=\left\{n_\mathrm{H}^t, T^t\right\}$ and the relative output
vector will be $\bar{y}=\left\{n_\mathrm{H}^{t+\Delta t},
T^{t+\Delta t}\right\}$, and consequently $n_x=n_y=2$. In this
example the number of inputs is equal to the number of outputs, but
this is not a strict requirement. In general, to avoid degeneracies
one should have $n_x\ge n_y$ (considering indipendent database parameters). 
The input parameters can be also
different from the output parameters, so in the previous example it
could be $\bar{y}=\left\{n_\mathrm{H}^{t+\Delta t},
n_\mathrm{H_2}^{t+\Delta t}\right\}$.

\begin{figure}
\begin{center}
\includegraphics[width=.45\textwidth]{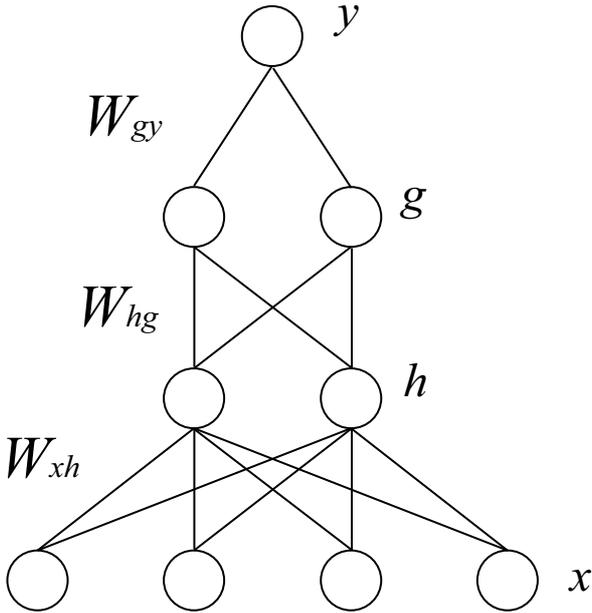}
\caption{The architecture of \textmd{MaNN}. Circles indicate
neurons, lines are connections (i.e. weights). Input units are
labeled with the symbol $\bar{x}$, hidden neurons with $\bar{h}$ if
{they} belong to the first layer, and $\bar{g}$ if
{they} belong to the second one; finally, output units are
indicated with $\bar{y}$. The matrix $W_{xh}$ represents the weights
between input and the first hidden units, $W_{hg}$ are the
connections between the first hidden units and the {second
hidden layer} and $W_{gy}$ connects the second layer of hidden units
and the output. A single output is shown. In this figure the forward
propagation is from lower neurons to the upper ones.
Back-propagation is from $\bar{y}$ to $\bar{x}$ neurons. Here
$n_x=4$, $n_h=n_g=2$ and $n_y=1$.}\label{annfig}
\end{center}
\end{figure}

There are two more layers in the \textmd{MaNN}, both are hidden and
generally with the same size, which in turn  depends on the
complexity of the task (for example, retrieving a single output
value can be obtained with two hidden layers of five elements, or
even less). The first layer is $\bar{h}$ with $n_h$ neurons and
$\bar{g}$ is the second one with $n_g$ units. Each layer is
connected with a matrix of weights. A $n_x \times n_h$ matrix named
$W_{xh}$ is situated between the input units and the first hidden
layer. Similarly, there {is} a $n_h \times n_g$ matrix named
$W_{hg}$ and, finally, $W_{gy}$, which is a $n_g \times n_y$ matrix.
The
 architecture of  \textmd{MaNN} that we have described here is sketched in Fig.
\ref{annfig}.

\subsection{The learning stage}\label{algorithm}

The whole learning process is divided into two different steps; the
first one is the so called \emph{training stage}, where
\textmd{MaNN} is trained with a randomly chosen set of data. In the
subsequent  \emph{test stage}  new, blank (i.e. never used before)
data are presented to \textmd{MaNN} to verify its behavior and the
quality of the training. If  \textmd{MaNN} retrieves the new data
with a small (user defined) error, than the training stage is
terminated. In case the error is still large, a new training stage
is required. The training stage can also be stopped if the absolute
error starts to increase again after an initial decrease
(\textit{early-stopping}). The aim is  to avoid the so called
\textit{over-training}:  a network is over-trained when it adapts
too strictly to the data fed during the supervised learning, so that
it cannot generalize the prediction to reproduce new cases of  the
test set of data, which contains data patterns that were not present
in the previous stage.

At the beginning of the training process, each weight is initialized
to a  small, randomly chosen,  value in the range $[-0.1,0.1]$. The weights
 change under the action of \textmd{MaNN}.  The  rate at which
\textmd{MaNN} changes its weights, otherwise known as the learning
rate $\eta$,  must be chosen carefully: for too high a value
\textmd{MaNN}  could  oscillate without reaching a stable
configuration, whereas for too small a value \textmd{MaNN} could
take a long time to reach the desired solution or, even worse, be
trapped in a false minimum of the error hyper-surface. We usually
choose $0.1\le\eta\le 0.9\,$.

The matrix of initial parameters  contains $N$ rows (the  initial
configurations of the input models); we randomly pick from this set
$N_L<N$ rows for the training stage, which we refer to as $L$, and
$N_T=N-N_L$ rows for the test stage, $T$.

A set $\bar{s}$  of initial parameters randomly picked up from $N_L$
is fed to the network as input vector $\bar{x}=\bar{s}$. The
activation of each node belonging to the first hidden layer is then
computed as
\begin{equation}\label{hideqn}
    \bar{h}=f\left(\bar{x}\otimes W_{xh}\right)\,,
\end{equation}
where $\bar{h}$ is the vector of hidden neurons,  $f$ is the sigmoid
function, $W_{xh}$ is the weight matrix between {the} input
and {the} first hidden neurons, and $\otimes$ indicates the
matrix product. If $\bar{c}=\bar{a}\otimes B$, where $\bar{a}$ is a
vector of $n$ elements and $B$ is a matrix $n\times m$, the size of
the vector $\bar{c}$ is $m$. The sigmoid function
$f\left(\bar{v}\right)$ has $\beta=1$. Then the following relation
holds
\begin{equation}
    \bar{g}=f\left(\bar{h}\otimes W_{hg}\right)\,,
\end{equation}
with obvious meaning of the symbols. Finally, the output vector of
\textmd{MaNN}  is given by
\begin{equation}\label{outeqn}
    \bar{y}=f\left(\bar{g}\otimes W_{gy}\right)\,.
\end{equation}
This first step  is named the \emph{forward-propagation}, because
the algorithm calculates the output of the network for a given input
vector $\bar{x}=\bar{s}$.

During the very first forward-propagation the output vector will
contain nearly random values, because the weights are randomly
generated. To improve upon it,  we  compute the error and
back-propagate it to change the weights accordingly.
The error at the   output layer is
\begin{equation}\label{deltay}
    \bar{\delta}_y=\left(\bar{t}-\bar{y}\right)\dot f\left(\bar{A}_y\right)=
    \left(\bar{t}-\bar{y}\right)\bar{y}(1-\bar{y})\,,
\end{equation}
where $\bar{A}_y$ is the vector of the activation functions for  the
output layer and the last equality is valid only with the activation
function $f\left(\bar{v}\right)$ with  $\beta=1$ {that} we
have adopted. The error $\bar{\delta}_y$ is propagated back to the
second hidden layer to become
\begin{equation}
    \bar{\delta}_g=\dot f\left(\bar{A}_g\right)
    \left[\bar{\delta}_y\otimes W_{gy}\right]=
    \bar{g}\left(1-\bar{g}\right)\left[\bar{\delta}_y\otimes W_{gy}\right]\,,
\end{equation}
with obvious meaning of all {the} symbols and expressions.
Finally the error is propagated back for the last time giving
\begin{equation}
    \bar{\delta}_h=\dot f\left(\bar{A}_h\right)
    \left[\bar{\delta}_g\otimes W_{hg}\right]=
    \bar{h}\left(1-\bar{h}\right)\left[\bar{\delta}_g\otimes W_{hg}\right]\,.
\end{equation}
Now that the error for each step between {the various} layers
has been computed, the changes of the synaptic weights are
\begin{eqnarray}
 \Delta W_{xh}&=&\eta \left[\bar{x}\otimes\bar{\delta}_h\right]+\alpha\Delta W_{xh}^o\,,\nonumber\\
 \Delta W_{hg}&=&\eta \left[\bar{h}\otimes\bar{\delta}_g\right]+\alpha\Delta W_{hg}^o\,,\nonumber\\
 \Delta W_{gy}&=&\eta \left[\bar{g}\otimes\bar{\delta}_y\right]+\alpha\Delta W_{gy}^o\,,\nonumber\\
\end{eqnarray}

The last terms at the r.h.s. of the above equations is called the
\emph{momentum} made by the product of a parameter $\alpha$ (to be
tuned) and the variation of the corresponding weights at the
previous step. $\Delta W_{xh}^o$ and the other similar terms
indicate the values of the weights at the previous training step.
The \emph{momentum} is introduced to force the procedure to avoid
false minima in the error hyper-surface.  This loop must be repeated
for each combination of {the} parameters in the set $L$.

To test the efficiency of {the} \textmd{MaNN} learning, the
forward propagation is then applied to all the variables belonging
to the set $L$.  The RMS of the test data set is
\begin{equation}\label{rms}
\mathrm{RMS}=\sqrt{\frac{\sum_{i=1}^{N_T}
\left(y_i-t_i\right)^2}{N_T-1}}\,,
\end{equation}
where $y_i$ is the output for the $i$-th test, $t_i$ is the $i$-th
correct test pattern and consequentially $y_i-t_i$ is the
\textmd{MaNN}'s error.

{The training test loop will be repeated until the MaNN's
learning reaches one of the following cases: (i) the RMS is lower
than a certain limit chosen by the user, (ii) the RMS does not
change for a sufficient number of loop cycles or, finally, (iii) the
early-stopping condition is verified.}

\begin{figure}
\begin{center}
\includegraphics[width=.45\textwidth]{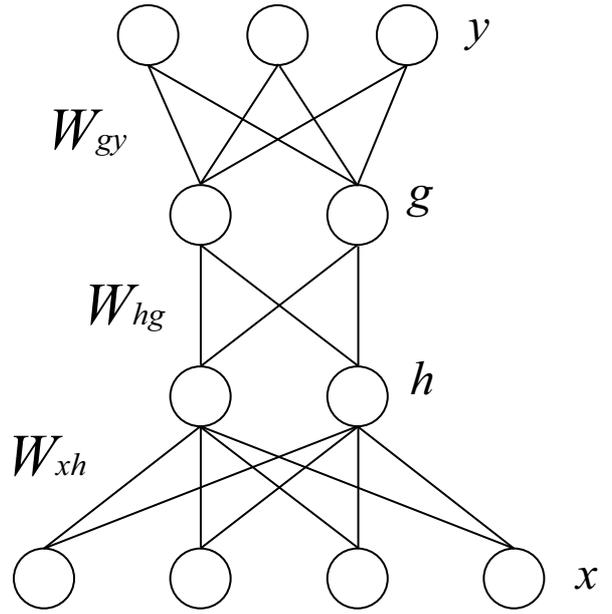}
\caption{The architecture of a double hidden layer ANN. Circles
indicate neurons, {while lines represent} the connections
{between the layers} (i.e. weights). {The} input units
are labeled with the symbol $\bar{x}$, {the} hidden neurons
with $\bar{h}$ if {they} belong to the first layer, and
{with} $\bar{g}$ if {they} belong to the second one;
{finally, the} output units are indicated with $\bar{y}$. The
matrix $W_{xh}$ represents the weights between {the} output
and the first hidden units, $W_{hg}$ are the connections between the
first hidden units and the second one and $W_{gy}$ connects the
second layer of hidden units and the output. Multiple outputs are
shown. In this figure the forward propagation is from {the}
lower neurons to the upper ones. {The} back-propagation is
from $\bar{y}$ to $\bar{x}$ neurons. Here $n_x=4$, $n_h=n_g=2$ and
$n_y=3$.}\label{ann2fig}
\end{center}
\end{figure}

\section{Training and testing MaNN}\label{testing}

\subsection{Input data from Robo}\label{data_from_robo}

As already stated, with \textsc{Robo} we {have calculated}
approximately 55000 models. {For a given set} of initial
conditions and physical parameters, each model describes the state
of an ISM element after $\Delta t = 10^4$ years of evolution.  This
time interval is short enough to secure that only small changes in
gas abundances have occurred (a sort of picture of the physical
state at given time), and long enough to be compatible with the
typical time step in large-scale dynamical simulations.

For the purposes of the present analysis, we limit ourselves to
track the following  parameters: temperature $T$, metallicity
$\tilde Z=10^{[\mathrm{Fe}/\mathrm{H}]}$\, \footnote{We {call
the attention of the reader on the definition} of metallicity
{that} we have adopted. {It is related} only to the
iron content in the standard spectroscopic notation and not to the
classical sum of the by mass abundances of all {the} species
heavier than $\mathrm{He}$, shortly indicated with $Z$ and
satisfying the {relation} $X+Y+Z=1$, where $X$ and $Y$ are
the by mass abundances of hydrogen and helium respectively.}, and
number densities  of $\mathrm H$, $\mathrm H_2$, $\mathrm H^+$ and
$\mathrm e^-$. This means that in addition to the temperature,
\textmd{MaNN} is designed here to follow just these four species,
whereas \textsc{Robo} can actually follow twenty-three species
including atoms, molecules and free electrons. The main motivation
for this limitation is that  in a large numerical dynamical
simulation it is more convenient in terms of computing time and
workload to follow in detail  only those species (and their
abundances)   that are known to play a key role in the real
evolution of the ISM (with particular attention to star formation
and energy feedback) and {to consider all the other species}
to remain nearly constant and small. In brief, the heavy elements
are important coolants {and} the same holds for the molecular
hydrogen, at least below $\sim 10^4$ K. The ionization fraction of
hydrogen is also important, because free electrons are involved in a
number of chemical reactions, and are the colliders that excite the
{coolants: this} is the reason why $n_\mathrm{H^+}$ and
$n_\mathrm{e^-}$ are considered. Moreover, the neutral
$n_\mathrm{H}$ is in the list to ensure the mass conservation while
changing {the} fractional abundances. Finally, the
temperature determines the kind of physical processes that are
important.

{The initial values of the number densities}. The initial
values of the number densities fall in three groups:  (i)
{the} elemental species with constant initial values, the
same for all {the models (namely $\mathrm{H^-}$,
$\mathrm{H_{2}^+}$, $\mathrm{He^+}$ and $\mathrm{He^{++}}$)}; (ii)
{the} elemental species whose initial values are derived from
other parameters {(namely $\textrm{He}$, all the deuteroids
and the metals, like for instance $\mathrm{C}$ and $\mathrm{O}$,
that depend on the choice {of the} total metallicity $\tilde
Z$)} {and, finally, (iii)} the elemental species with free
initial conditions {(namely $\mathrm{H}$, $\mathrm{H_2}$,
$\mathrm{H^+}$, and $e^{-}$).}

\noindent\textsc{Hydrogen group: $\mathrm{H}$, $\mathrm{H^+}$,
$\mathrm{H^-}$, $\mathrm{H_2}$} {and}
\textsc{$\mathrm{H_2^+}$}. {The initial values of the species
$\mathrm{H^-}$ and $\mathrm{H_2^+}$ are $n_{\mathrm{H}^{-}}=n_{\mathrm{H}_2^{+}}=10^{-10}$cm$^{-3}$
\cite{Prieto2008}, while the other three hydrogen species have free
initial values.}

\noindent\textsc{Deuterium group: $\mathrm{D}$, $\mathrm {D^+}$,
$\mathrm {D^-}$, $\mathrm{D_2}$, $\mathrm{HD}$} {and}
\textsc{$\mathrm{HD^+}$}. The numerical densities of {the}
deuteroids are calculated from their hydrogenoid counterparts. For
{the} single atom species we have
$n_\mathrm{D}=f_\mathrm{D}\,n_\mathrm{H}$,
$n_\mathrm{D^+}=f_\mathrm{D}\,n_\mathrm{H^+}$ {and}
$n_\mathrm{D^-}=f_\mathrm{D}\,n_\mathrm{H^-}$, where
$f_\mathrm{D}=n_\mathrm{D}/n_\mathrm{H}$. For the molecules, we can
consider the ratio $f_\mathrm{D}$ as the probability of finding an
atom of deuterium in a {population of hydrogen-deuterium
atoms.} This assumption allows us to calculate {the}
$\mathrm{HD}$, $\mathrm D_2$ and $\mathrm{HD^+}$ number densities as
a joint probability; for $\mathrm{HD}$ and $\mathrm{HD}^+$ we have
$n_\mathrm{HD}=f_\mathrm{D}\,n_\mathrm{H_2}$ and
$n_\mathrm{HD^+}=f_\mathrm{D}\,n_\mathrm{H_2^+}$, but for $\mathrm
D_2$ is $n_\mathrm{D_2}=f_\mathrm{D}^2\,n_\mathrm{H_2}$ as the
probability of finding two deuterium atoms is $f_\mathrm{D}^2$. This
is valid as long as $f_\mathrm{D}\ll1$.

\noindent\textsc{Helium group:  $\mathrm{He}$, $\mathrm{He}^+$,
$\mathrm{He}^{++}$}. {The ratio $\mathrm{n_{He}/n_H}$ is set
to 0.08, thus allowing to the initial value of $\mathrm{n_{He}}$ to
vary} accordingly to the initial value for $\mathrm{n_{H}}$. The
initial number densities of the species $\mathrm{He}^+$,
$\mathrm{He^{++}}$ are both set equal zero and kept constant in all
{the} models.

\noindent\textsc{Metals group:  $\mathrm{C}$,  $\mathrm C^+$, O,
$\mathrm O^+$, $\mathrm{Si}$, $\mathrm{Si}^+$} {and}
\textsc{$\mathrm{Fe}$, $\mathrm{Fe}^+$}.  The $\mathrm{Fe}$ number
density of the ISM is
\begin{equation}
n_\mathrm{Fe}=n_\mathrm{H}\cdot \mathrm{dex}\left\{[\mathrm{Fe/H}]+
\log\left(\frac{n_\mathrm{Fe}}{n_\mathrm{H}}\right)_\odot\right\}\,,
\end{equation}
where $\left(n_\mathrm{Fe}/n_\mathrm{H}\right)_\odot$ is the
iron-hydrogen ratio for the Sun. To retrieve the number density of a
given metal X we use $n_\mathrm{X}=n_\mathrm{Fe}\cdot f_\mathrm{X}$,
where $f_\mathrm{X}$ is the metal-iron number density ratio in the
Sun.

The list of {the} species whose initial number densities are
kept constant in all {the} models of the ISM is given in
Table \ref{fixparam}. The chemical composition of the ISM is typically
primordial with the by mass abundances of hydrogen $X= 0.76$, and
helium $Y=0.24$ and all metals $Z \simeq 0$ . {The helium to
hydrogen number density ratio corresponding to this primordial by
mass abundances is $n_{\textrm{He}}/n_{\textrm{H}} \simeq 0.08$.}
The {adopted} cosmological ratio for {the} deuterium
is $f_\textrm{D} = n_\textrm{D}/n_\textrm{H} \simeq 10^{-5}$. 
With the aid of these numbers and the above prescriptions
we get the number density ratios listed in Table \ref{fixparm} and
the initial values of the number density in turn.

 {Domains of the physical variables}.  In the present
version of \textmd{MaNN} the variables in question span the
following ranges: $\tilde Z=[10^{-12},10]$ for the metallicity,
$n_\mathrm{H}=n_\mathrm{H^+}=n_\mathrm{e^-}=n_\mathrm{H_2}=[10^{-12},10^3]\
\mathrm{cm^{-3}}$ for the numerical densities, and $T=[10,10^7]\
\mathrm K$ for the temperature.

{Role of the dust}. In this study, we neglect the presence of
dust: in other words, the \emph{dust density} is kept constantly
equal to zero. This is because the current version of the NB-TSPH
code \textsc{EvoL} does not include the treatment of the dust
component of the ISM. We plan to include silicates and carbonaceous
dust grains both in \textsc{EvoL} and \textmd{MaNN}.

{Other parameters}.  In addition to this, we do not consider
the contribution by the \emph{Cosmic Rays field}, {the}
\emph{background heating}, and {the} \emph{UV ionizing
field}. {All of these processes, however, can be described by
\textsc{Robo} \citep[see][\, for more details]{Grassi10}.}

The CMB temperature is set to $2.725\ \mathrm{K}$, the observed
present value \citep{Fixsen2009} {and is kept constant in all
the models}. In {the} case of a cosmological use of
\textmd{MaNN} {one should include a} redshift-dependent
variation of the CMB temperature {for a correct description
of the CMB effects.}

The parameters we have chosen form a $\mathbb{R}^6$ {space:
the} input space of the \textmd{MaNN}. The large number of input
models secures that the space of  physical parameters is smoothly
mapped.

Finally, in addition to the six-dimensional input vector, another
input unit is dedicated to the so called \emph{bias} which is always
set to $x_0=-1$; it gives the threshold $\theta$ in the eqn.
(\ref{activation}).

\renewcommand{\arraystretch}{1.1}
\begin{table} \label{fixparm}
\begin{center}
\caption{{Initial values for the number densities of the
hydrogen and helium elemental species and the deuteroids. They are
either fixed to a constant value or based upon the numerical
abundance of one of the \textit{free} hydrogen species
$n_{\textrm{H}}$, $n_{\textrm{H}_{\textrm{2}}}$ and
$n_{\textrm{H}^{+}}$ via the $f_{\textrm{D}}$ factor. Since $\mathrm{H^-}$
and $\mathrm{H_2^+}$ are constant, then also $\mathrm{D^-}$ and
$\mathrm{HD^+}$ are fixed. Values are indicated as $a(b)=a\times
10^b$. See also the text for details.}} \label{fixparam}
\vspace{1mm}
\begin{tabular*}{.32\textwidth}{l l| l l}
\hline
$\mathrm{H^-}$     &$1.0 (-12)$   &$\mathrm{D}$    &$f_{\textrm{D}}$n$_{\mathrm{H}}$\\
$\mathrm{H_2^+}$   &$1.0 (-12)$  &$\mathrm{D^+}$  &$f_{\textrm{D}}$n$_{\mathrm{H}^{+}}$\\
$\mathrm{He}$      &$0.8 (-1) $ n$_{\mathrm{H}}$  &$\mathrm{D^-}$  &$f_{\textrm{D}}$n$_{\mathrm{H^{-}}}$\\
$\mathrm{He^+}$    &$0.0 (+0) $   &$\mathrm{HD}$   &$f_{\textrm{D}}$n$_{\mathrm{H_2}}$\\
$\mathrm{He^{++}}$ &$0.0 (+0) $   &$\mathrm{HD^+}$ &$f_{\textrm{D}}$n$_{\mathrm{H_{2}^{+}}}$\\
                    &               &$\mathrm{D_2}$&$f_{\textrm{D}}^{2}$n$_{\mathrm{H_2}}$\\
\hline
\end{tabular*}
\end{center}
\end{table}
\renewcommand{\arraystretch}{1}

\begin{figure*}
\begin{center}$
\begin{array}{cc}
\includegraphics[width=.45\textwidth]{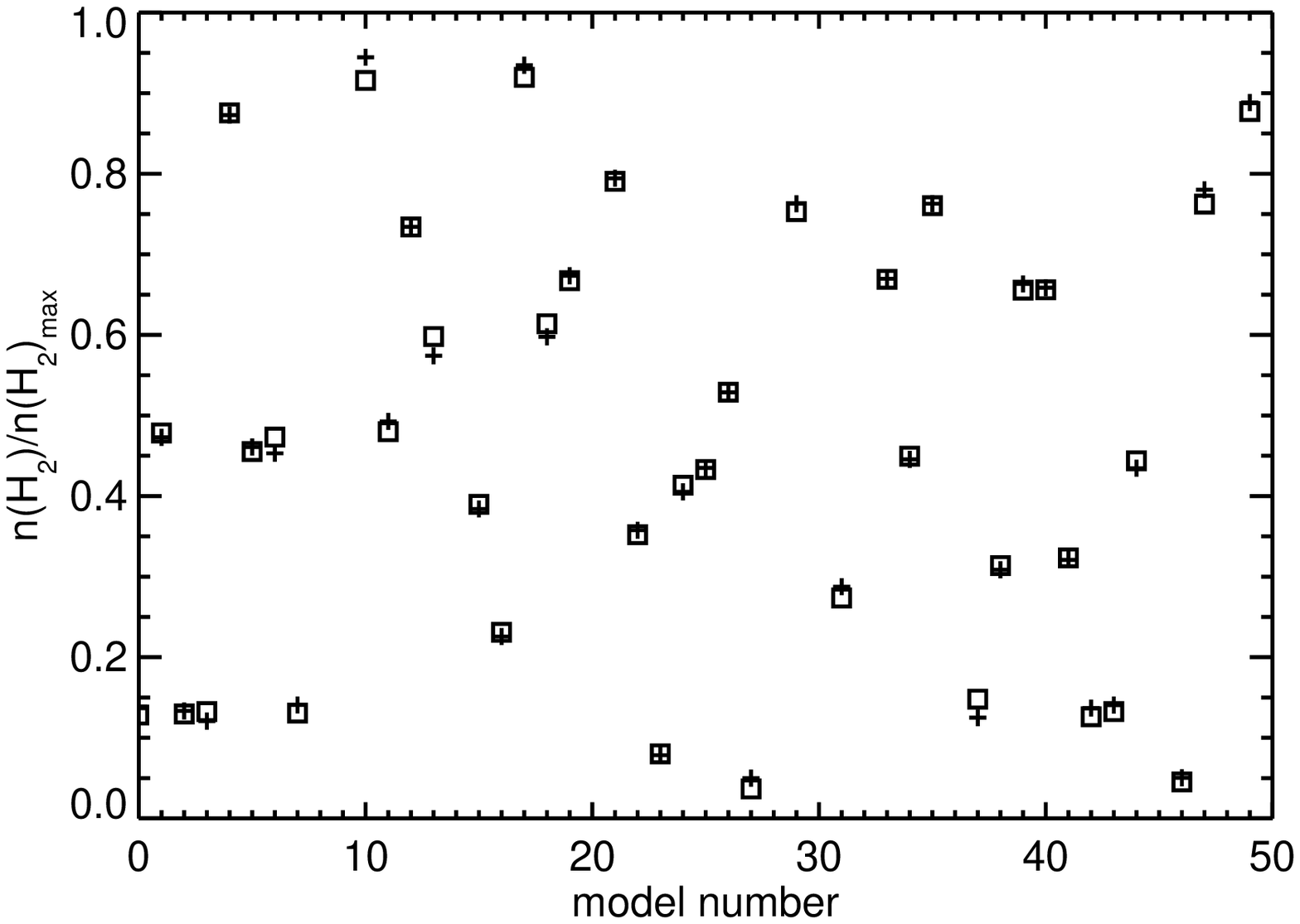}&
\includegraphics[width=.45\textwidth]{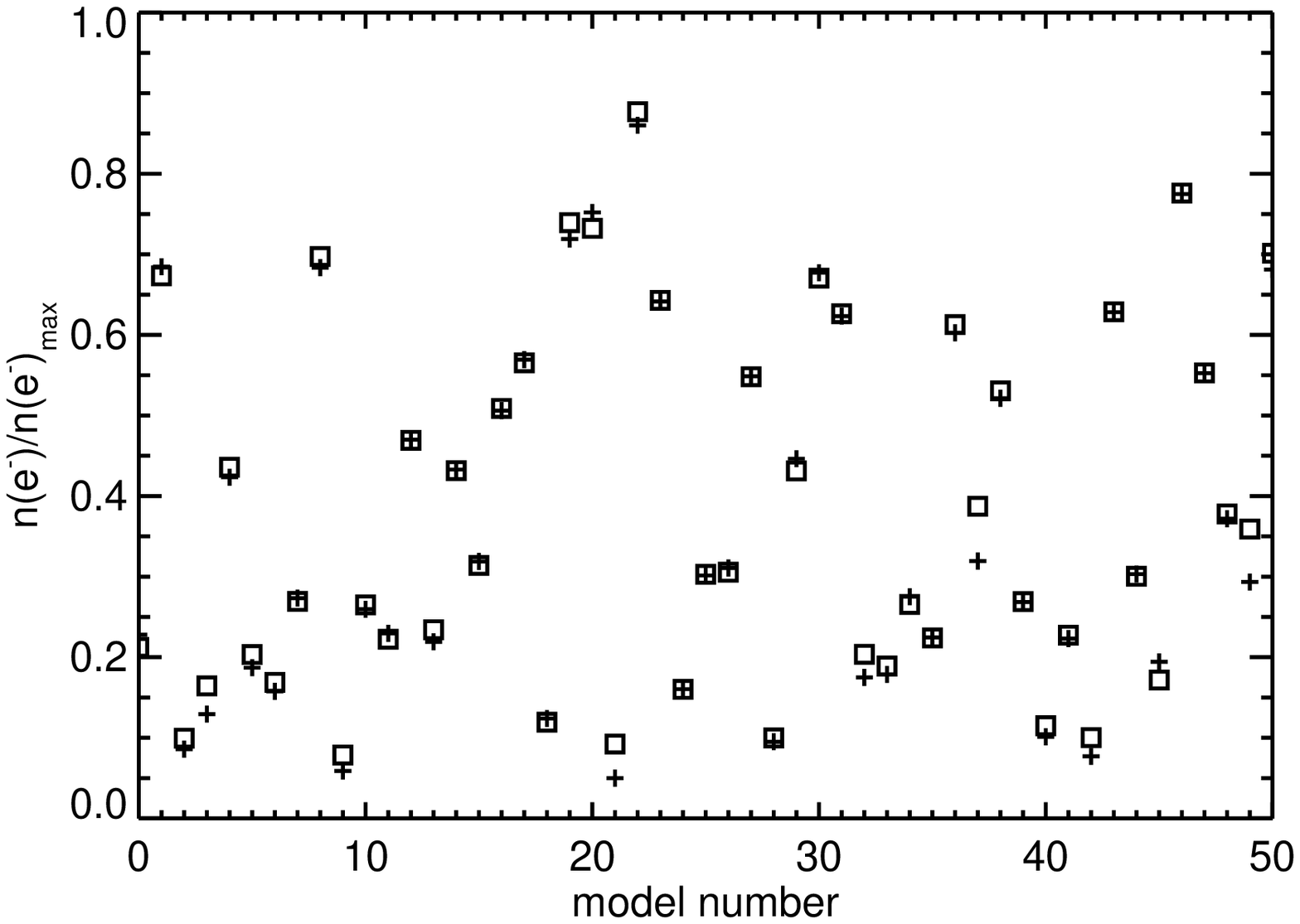}
\end{array}$\caption{Comparison between normalized data from
\textmd{MaNN} (cross) and \textsc{Robo} (square), for molecular
hydrogen density (left panel) and free electrons (right panel). We
show only 50 randomly picked models. Similar results for the other
outputs.}\label{res}\end{center}
\end{figure*}

\subsection{Data normalization}
All the input data are normalized. Given a generic input model
contained in the vector $\bar{s}$ of dimensions $n_s $, first we
take the logarithm (base 10) of its components (all of which are
positive quantities) and replace it with a new vector $\bar{s}' $ so
defined
\begin{equation}
 \bar{s}'=\log(\bar{s}+\bar{\epsilon}) \nonumber
\end{equation}
where $\bar{\epsilon}$ is a vector of the same length as $\bar{s}$
made of constant, small positive quantities  here introduced to
avoid infinities in the logarithms (or zeros in the original
quantities). In all {the cases, all the} components of
$\bar{\epsilon}$ are $\epsilon=10^{-12}$. Then we introduce the
matrix $L$ containing all the $\bar{s}'$ vectors. The rows of
{the} matrix $L$ are input models, one row for each of them,
whereas the columns of $L$ are the $n_s$ elements of the $\bar{s}'$.
Finally we calculate the vector $\bar{x}$

\begin{equation}\label{norm}
    \bar{x}=\frac{\left(\bar{s}'-\langle\bar{s}'\rangle\right) -\mathrm{MIN}\left[L\right]}
    {\mathrm{MAX}\left[L\right]-
    \mathrm{MIN}\left[L\right]}\left(n_+-n_-\right)+n_-\,,
\end{equation}

\noindent where  $\mathrm{MAX}$ and $\mathrm{MIN}$ are functions
that retrieve the $\bar{s}$-sized maximum and minimum value from
their argument, $\langle\bar{s}'\rangle$ is a vector with size 
equal to the number of input parameters. The last term is introduced
to have a zero-mean sample of data.
the mean of $\bar{s}'$. 
The vector $\bar{x}$ is the so-called input pattern
of   \textmd{MaNN}. The quantities  $n_+=0.95$ and $n_-=0.05$ are
the upper and the lower bounds of the normalization interval. In
other words, for each component of the vector $\bar{s}'$ we have
searched in the corresponding columns of the matrix $L$ the minimum
and maximum values with which we have normalized the $\bar{s}'$
element.

To avoid notation misunderstanding, we recall that
$s_i\in\bar{s}\subset L$ and $x_i\in\bar x$ with $x_i$ the $i$-th
neuron belonging to the input layer $\bar x$. When a pattern is used
as input, we have $\bar x \equiv \bar{s}'$.

Similar procedure is adopted for the output {values of
\textmd{MaNN}.} Let $\bar{t}$ be the vector of the original output
models, $\bar{t}'$ the vector
$\bar{t}'=\log(\bar{t}+\bar{\epsilon})$ {and} $T$ the matrix
containing all the $\bar{t}'$ {vectors (the analog of $L$):}
the vector of normalized output values $\bar{t}_n$ is

\begin{equation}\label{normt}
    \bar{t}_n=\frac{\left(\bar{t}'-\langle\bar{t}'\rangle\right)-\mathrm{MIN}\left[T\right]}
    {\mathrm{MAX}\left[T\right]-
    \mathrm{MIN}\left[T\right]}\left(n_+-n_-\right)+n_-\,,
\end{equation}
with the same meaning  of eqn. (\ref{norm}). The notation here is
$t_i\in\bar{t}\subset{T}$.

\subsection{Compression of data?}\label{preprocessing}

To reduce the complexity of the problem, one can try to compress the
input data, i.e. to reduce the number of variables and to obtain a
new initial set easy-to-handle. A popular technique to reduce the
number of dimensions of a data sample is the \textit{principal
component analysis} (PCA). The PCA is a transformation between the
original data space and a new space (named PCA space). The aim of
this method is to build this $\mathbb{R}^n$-dimensional subspace
(with $n$ less than {the} number of dimensions of the
original space) so that its new PCA-variables are uncorrelated. In
this way, only the most statistically significant parameters (i.e.,
the new variables that have the highest percentage of the total
variance) are taken into consideration, instead of using the
original parameters. PCA is useful when the data set is reduced to
two or three modes. This is because two matrix products are needed,
one to transform the input from the physical parameters space to the
PCA space, and one to transform the output from the PCA space to the
physical parameters space. These two operations have a
non-negligible computational cost. In general, compressing the data
with the PCA may improve the learning process of \textmd{MaNN} and
its ability in reproducing the original data.

Analyzing the issue, we soon found  that in our case the PCA method
is not useful, and using the original data  yields  better results.
Table \ref{PCA} contains the  results of the PCA analysis. It
{is} easy to understand  why reducing the number of
parameters does not improve the situation: five modes cover
approximately  87\% of the total variance, and with the sixth mode
the coverage reaches  94\%. This means that the reduction of the
number of parameters to obtain a $\mathbb{R}^2$ or $\mathbb{R}^3$
space is not possible. Note that only the PCA on the output data is
shown, since the input data are random and they generate six modes
with the same variance.

\begin{table}
\begin{center}
\caption{Results of the PCA analysis on the output data.}\label{PCA}
 \vspace{1mm}
\begin{tabular*}{.30\textwidth}{c r@{.}l r@{.}l}
\hline
mode & \multicolumn{2}{l}{Eigenvalue} & \multicolumn{2}{l}{Variance (\%)}\\
\hline
$1$ &$0$&$14585$ &$36$&$9657$\\
$2$ &$0$&$07454$ &$18$&$8922$\\
$3$ &$0$&$07006$ &$17$&$7580$\\
$4$ &$0$&$05606$ &$14$&$2073$\\
$5$ &$0$&$02871$  &$7$&$2760$\\
$6$ &$0$&$01934$  &$4$&$9008$\\
\hline
\end{tabular*}
\end{center}
\end{table}

\subsection{Single or multiple ANNs for  output configurations}\label{architecture}

There are different ways in which the ANN algorithm can {be}
used in practice to extract {the results that} we need from a
given initial set of data. {Once} given {an} input
pattern one may use a single ANN with multiple outputs (one for each
quantity to evaluate). {This} situation is illustrated in
Fig. \ref{ann2fig}. We will refer to it as case A or single ANN.
Alternatively, given the same input pattern, one
may want to set up a manyfold of ANNs, each of which with a single
output. {This} situation for each ANN {is
corresponding} to the one shown in Fig. \ref{annfig}. We will refer
to it as case B and the whole algorithm is named multiple ANN (from
which the acronym \textmd{MaNN} we have already introduced). Note
that this difference would be unrelevant if the \textmd{MaNN} does
not include hidden layers, because in that case the outputs would be
uncorrelated. However, when one or more layers are added, the
scenario changes. As Eqn. (\ref{outeqn}) shows, the state of an
output neuron depends on the weights that connect it to the previous
layer. If an hidden layer is present, then all the output neurons
rely on every hidden neurons (albeit with different weights). These
neurons are at the same time linked with a weight matrix to the
input layer (or with another hidden layer). In this way, two output
neurons share some weights. This is crucial when the
back-propagation error is applied, because different output neurons
interfere modifying the same weight (i.e. the shared one).

There is another difference worth being pointed out. Input and
output data patterns are not physically uncorrelated {because
the latter results} from the former via the \textsc{Robo} model. The
architecture of case A by somehow forces an artificial correlation
between input and output data patterns other than the natural one
established by the underlaying ISM physics. This may originate from
the back-propagation procedure simultaneously changing all
{the} weights. Alternatively, case B does not contain this
undesired effect. The input pattern separately determines each
output value.

As a matter of fact, using the case A, during the learning phase the
way the error decreased was not satisfactory. So we put case A aside
and turned to case B. The set up of  \textmd{MaNN} includes six
different networks, one for each output parameter. It also contains
a double hidden layer network  because other schemes (single hidden
layer or no hidden layers at all) did not give good results.

\subsection{Training results}\label{eps_test}

We apply \textmd{MaNN} to our database of ISM models obtaining
different performances for different network parameters. The best
cross-correlation results are obtained for $n_h=20$ and $n_g=200$.
Using a smaller number of hidden neurons yields unsatisfactory
results, as well as with $n_h>n_g$. Bad results are obtained when
the RMS decrease is slow, when the error is high ($>10^{-3}$) just
before applying the early-stopping technique, or, finally, when the
RMS curve appears to be noisy (huge RMS oscillations). The learning
rate has been set to $\eta=0.5$, which appears to be the best
compromise between learning speed and accuracy.

Results obtained with the choice of these parameters are shown in
Fig. \ref{res}, with crosses indicating the data retrieved with
\textmd{MaNN} and squares {used to show the values taken}
from the database produced with \textsc{Robo}. The two panels show
molecular hydrogen and free electrons abundances, both normalized
following eqn. (\ref{normt}). We only show 50 randomly picked
models. The other outputs ($n_\mathrm{H}$ and $n_\mathrm{H^+}$) are
not displayed here since they have a similar aspect.

As the figure shows, the error on the test set is relatively low.
\textmd{MaNN} retrieves the data produced with \textsc{Robo} fairly
well. It is therefore possible to use it in the context of complex
dynamical simulations.

\begin{figure*}
\begin{center}
\includegraphics[width=14cm,height=10cm]{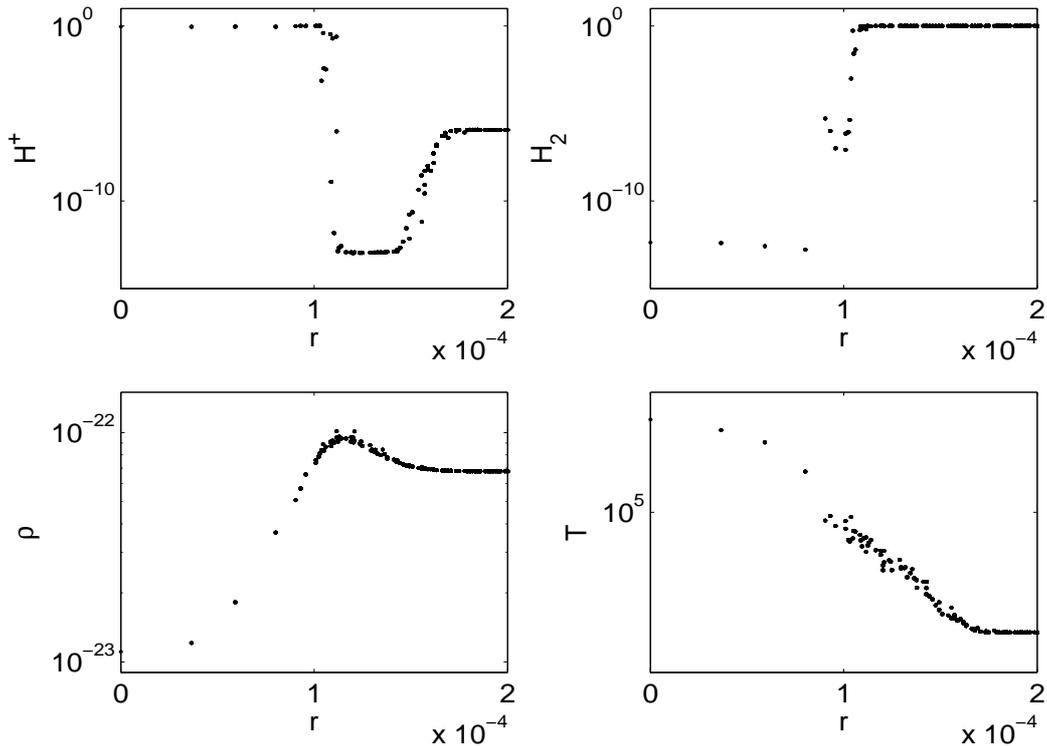}
\caption{Sedov-Taylor OB-wind test with chemical evolution.
Left to right, top to bottom: radial profiles of $\mathrm{H^+}$
and $\mathrm{H_2}$ mass fractions, $\rho$, and T, at $t=0.5$ Myr.
Plotted are all particles within the central 200 pc.}
\label{profs}
\end{center}
\end{figure*}

\begin{figure*}
\begin{center}
\includegraphics[width=14cm,height=5cm]{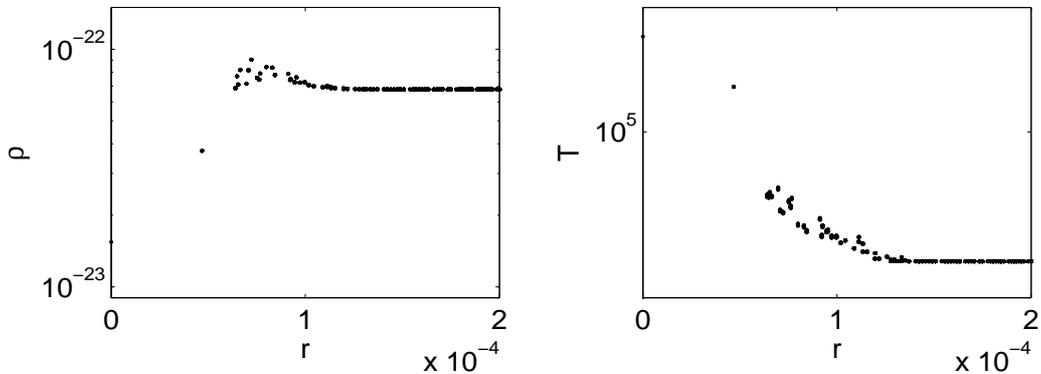}
\caption{Sedov-taylor OB-wind  test without chemical evolution.
Left to right: radial profiles of $\rho$ and T, at $t=0.5$ Myr.
Plotted are all particles within the central 200 pc.}\label{profso}\end{center}
\end{figure*}

\section{MaNN interfaced with  NB-TSPH simulations}\label{application}

To evaluate the performance of  \textmd{MaNN} in real large -scale
numerical simulations we {implemented} it in the NB-TSPH code
\textsc{EvoL}, and run a ``realistic'' version of the ideal
Sedov-Taylor problem, which is known to be a rather demanding test.

\textmd{MaNN} tracks the evolution of the relative abundances of
$\mathrm{H}, \mathrm{H^+}, \mathrm{H_2}$ and $\mathrm{e^-}$ within
each gas particle of the SPH simulation. The \textmd{MaNN} routine
is called by each active gas particle during the computation of its
radiative cooling \citep[which is in turn obtained from look-up
tables, see][\, for details]{Merlin2007}. Since the time interval
$\Delta t=10^4$ yr in between the initial and the final stage of any
model in \textmd{MaNN} is usually smaller than the typical dynamical
time step of a gas particle, the routine is called several times
until the dynamical time step is covered. If the dynamical time step
is not an integer multiple of the \textmd{MaNN} time interval, an
interpolation is applied to the results of {the} last two
calls of \textmd{MaNN}. During the whole procedure, all the physical
parameters but the afore mentioned relative chemical abundances and
the temperature (namely, total density, metallicity and mechanical
heating rate) are kept constant.

Due to the detailed description of the chemical properties of each
particle, the adiabatic index $\gamma$ in the equation of state
governing the gas  pressure  $P=(\gamma-1) \rho u$  varies as a
function of the relative abundances (see \citet{Grassi10}). This will immediately reflect on the relationship
between {the} internal energy and {the} temperature,
and on the equation of {the} motion itself. Therefore, we
expect an important change in the dynamical evolution of the whole
system as compared to the case in which these chemical effects are
ignored.

\begin{figure*}
\begin{center}$
\begin{array}{cc}
\includegraphics[width=15cm,height=6.5cm]{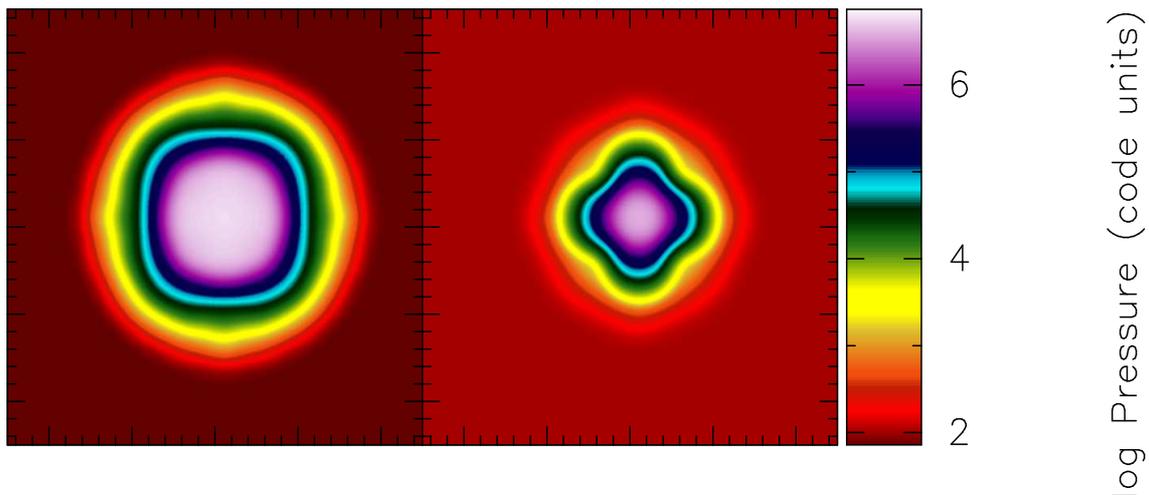}
\end{array}$\caption{Sedov-taylor OB-wind test. Pressure field in a
central slice of side $l=500$ pc, with (left) and without (right)
chemical evolution.}\label{pres}\end{center}
\end{figure*}

The test run is set up as follows. A uniform grid of $45^3$ equal
mass gas particles is displaced inside a periodic box of size $l=1$
kpc. The mass of each mass element is $\simeq 10^4$ M$_{\odot}$, so
the initial uniform density is $\rho \simeq 6.78 \times 10^{-23}$
g/cm$^3$. The initial temperature is $200$ K. The initial chemical
mass fractions of each gaseous element are $[\mathrm H^+] =
10^{-20}$, $[\mathrm H_2] = 0.99$, $\mathrm{H} = 1 -
([\mathrm{H_2}]+[\mathrm{H^+}])$. These initial conditions resemble
a dense interstellar cloud of cold molecular material.

At time $t=0$, the temperature of the central particle is
instantaneously raised up to $T = 2\times 10^7$ K, a temperature
similar to that of the gas shocked by the winds ejected by O-B young
stars. We name this physical case  as the \emph{Sedov-Taylor
OB-wind}. At the same time, the chemical composition of the particle
is changed to $[\mathrm H_2] = 10^{-20}$, $[\mathrm H^+] = 0.99$,
$\mathrm{H} = 1 - ([\mathrm{H_2}]+[\mathrm{H^+}])$. The temperature
of the central particle is maintained constant, to mimic the
continuous energy injection by the central star, and the system is
let  evolve freely under the action of hydrodynamic interactions and
chemical reactions. We also include  thermal conduction and
radiative cooling that are usually neglected  in standard
Sedov-Taylor explosion tests \citep[see][\, for more
details]{Merlin2010}.

A region of hot material expands in the surroundings of the central
particle. This is due to a spurious numerical  conduction which,
however, causes a numerical  smoothing of the initial condition and
favors the correct  developing of the subsequent evolutionary
stages. Then, a shock wave forms and moves outwards radially. Note
that the continuous energy release from the central particle and the
effect of  radiative cooling, almost prevent the inward motion of
shocked particles like in  standard tests of the Sedov-Taylor
explosion.

In Fig. \ref{profs} we show the state of the system after $\simeq
0.5$ Myr of evolution. The bottom panels display   the density and
temperature profiles, and the top panels the abundances of $\mathrm
H^+$ and $\mathrm H_2$.  In the central hot and less dense  region,
the molecular hydrogen gets completely ionized (the abundance of
$\mathrm H_2$ falls to very low values), due to the high
temperatures and low densities. The less dense cavity is surrounded
by a layer of over-dense material (the shock layer) which in turn is
surrounded by another  layer with  the initial density. The
over-dense front is moving outwards. In the   pre-shock region, near
the shock layer, the density increase  causes the $\mathrm{H^+}$ to
recombine into neutral hydrogen. These results fairly agree with the
theoretical predictions.

To strengthen the above conclusion, {we performed} another
simulation of the Sedov-Taylor OB-wind, in which {the effects
of the chemical composition and its changes on the gas particles are
ignored} (by the way this is exactly what is assumed in most if not
all {the} cosmological simulations).  In practice, all
{the} gas particles are assumed to be composed of solely
neutral H, so that the adiabatic index $\gamma$ is fixed and equal
to 5/3. the results are shown in Fig. \ref{profso}, in which the
density and temperature profiles are displayed at the {same}
age of the previous case {presented in Fig. \ref{profs}}.
Clearly, the dynamical evolution of the system is substantially
different. The over-dense front moves at significantly lower speed.
In particular, it is clear how the variable chemical composition
ensures a higher pressure in the central region containing ionized
particles, with respect to the outer region containing molecules.
This causes the faster expansion of the central bubble with respect
to present case. This is also supported by the pressure fields in
the two cases shown in Fig. \ref{pres}. The left panel shows the
pressure {field} of the first case with evolving chemical
composition of the gas particles, whereas the right panel shows the
same but for constant chemical composition. The color code shows the
value of the pressure in code units. Both pressure fields are
measured at the same age of 0.5 Myr and on the surface of an
arbitrary plane passing through the center of the computational box.
There are a number of interesting features to note: (i) The pressure
in the outskirts is lower in the case with chemical evolution, where
{the adiabatic index is} $\gamma=7/5$ for the bi-atomic
molecule $\mathrm H_2$. At the same time, the pressure in the
central region of the case with chemical evolution is higher by
nearly a factor of two because the number of free particles (or
equivalently  the internal energy corresponding to a given
temperature) is nearly doubled. (ii) The continuous energy release
from the central particle, plus the action of the radiative cooling,
almost prevent the inward motion of shocked particles (conversely to
what happens in the standard Sedov-Taylor explosion test). Instead,
they are compressed in a thin layer, which moves outwards under the
action of the central source of pressure. For this reason, we are
interested in the state of the system shortly after the beginning of
the energy release, before the shock layer is pushed too far away
from the central region. However, at that moment only a relatively
small number of particles have already felt a strong hydrodynamical
interaction. This explains the geometrical features that can be seen
in Figs. \ref{pres}. They are due to the initial regular
displacement of {the} particles, whose effects are still to
be smeared out by hydrodynamics interactions. Because the detailed
chemical treatment implies a larger pressure gradient and,
consequently, a faster expansion of the central bubble, after a
given $\Delta t$ the simulation without the inclusion of
\textmd{MaNN} shows a stronger geometrical spurious asymmetry with
respect to the ``standard'' case: this is simply because it is at a
more primordial stage of its evolution.

\section{Summary and conclusions}\label{conclusions}

We presented \textmd{MaNN}, an artificial neural network aimed at
providing a light and versatile interface between  databases of
models of the ISM and numerical hydrodynamical codes. {In our
specific case the ISM models are the outputs} of the \textsc{Robo}
{code} developed by \citet{Grassi10} and {the} Padova
NB-TSPH code \textsc{EvoL} developed by \citet{Merlin2010}.

{With the aid of \textsc{Robo}} we have calculated a large
database of ISM models (55000 in total). {In these models,}
given a set of initial conditions, the thermodynamical and chemical
evolution of the ISM is followed during a  given time interval to
get the new physical state of the ISM. A large volume of the
hyper-space o f the physical parameters and their initial conditions
is explored to get an ensemble of \textsc{Robo} models. For the aims
of this first study, the ISM models are calculated switching off the
dust, the external heating, and the cosmological evolution. They can
be {restored} at any time.

The above database is  fed to \textmd{MaNN} with some
simplifications. First of all, a \textmd{MaNN} model is made of two
strings (vectors)  of quantities: the initial condition for the key
parameters (input vector) {and} the results for the
quantities describing the final state after a certain time interval
of evolution (output vector). Not necessarily they must be identical
in dimension and type of physical quantities. The first step to
undertake {is} to train \textmd{MaNN} to reproduce the
results obtained with \textsc{Robo}. The technique in use is the
back-propagation algorithm. Once the training stage is completed and
the architecture of \textmd{MaNN} is fixed,  \textmd{MaNN} is able
to mimic the results of  \textsc{Robo} with a great degree of
accuracy. The great advantage of all {of} this is that in
large cosmological or galactic {simulations} the
thermodynamical and chemical properties of all {the} gas
particles can be determined at each time step with great accuracy in
practice at no {computational} cost.

Finally, we have implemented \textmd{MaNN} into \textsc{EvoL} and
run the Sedov-Taylor OB-wind test,   i.e. the evolution of a dense
and cold molecular cloud during the release of a high amount of
energy in its central region. The results are fairly good, showing a
nice agreement with the theoretical expectations: while a shock wave
develops and moves outwards, a hot and ionized region forms in the
central region. Taking into account the detailed chemical evolution
greatly affects the thermo-dynamical evolution of the system.
Similar effects {should} be expected in realistic NB-TSPH
simulations of cosmological or galactic large scale structures.

In conclusion, \textmd{MaNN} proves to be a practical,
easy-to-handle and robust method to include the thermal and chemical
properties of the ISM in NB-TSPH simulations.

\begin{acknowledgements}

T. Grassi is grateful to Dr. F. Combes for the kind hospitality at
the Observatoire de Paris - LERMA under EARA grants where part of
the work has been developed and the many stimulating discussions.
\end{acknowledgements}

\bibliographystyle{apj}
\bibliography{mybib}

\begin{thebibliography}{26}
\expandafter\ifx\csname natexlab\endcsname\relax\def\natexlab#1{#1}\fi

\bibitem[{{Bakes} \& {Tielens}(1994)}]{BakesTielens94}
{Bakes}, E.~L.~O. \& {Tielens}, A.~G.~G.~M. 1994, \apj, 427, 822

\bibitem[{{Carpenter} \& {Grossberg}(1987)}]{CarpenterGrossberg87b}
{Carpenter}, G.~A. \& {Grossberg}, S. 1987, \ao, 26, 4919

\bibitem[{Carpenter \& Grossberg(1987)}]{CarpenterGrossberg87a}
Carpenter, G.~A. \& Grossberg, S. 1987, Comput. Vision Graph. Image Process.,
  37, 54

\bibitem[{{Carpenter} \& {Grossberg}(1990)}]{CarpenterGrossberg90}
{Carpenter}, G.~A. \& {Grossberg}, S. 1990, Neural Networks, 3, 129

\bibitem[{{Cazaux} \& {Spaans}(2009)}]{CazauxSpaans09}
{Cazaux}, S. \& {Spaans}, M. 2009, \aap, 496, 365

\bibitem[{{Cen}(1992)}]{Cen92}
{Cen}, R. 1992, \apjs, 78, 341

\bibitem[{{Draine} \& {Salpeter}(1979)}]{DraineSalpeter79a}
{Draine}, B.~T. \& {Salpeter}, E.~E. 1979, \apj, 231, 77

\bibitem[{{Dwek}(1998)}]{Dwek98}
{Dwek}, E. 1998, \apj, 501, 643

\bibitem[{{Fixsen}(2009)}]{Fixsen2009}
{Fixsen}, D.~J. 2009, \apj, 707, 916

\bibitem[{{Galli} \& {Palla}(1998)}]{GalliPalla98}
{Galli}, D. \& {Palla}, F. 1998, \aap, 335, 403

\bibitem[{{Glover} \& {Jappsen}(2007)}]{GloverJappsen2007}
{Glover}, S.~C.~O. \& {Jappsen}, A. 2007, \apj, 666, 1

\bibitem[{{Grassi} {et~al.}(2010){Grassi}, {Krstic}, {Merlin}, {Buonomo},
  {Piovan}, \& {Chiosi}}]{Grassi10}
{Grassi}, T., {Krstic}, P., {Merlin}, E., {Buonomo}, U., {Piovan}, L., \&
  {Chiosi}, C. 2010, ArXiv e-prints 1012.1142

\bibitem[{{Hebb}(1949)}]{Hebb49}
{Hebb}, D.~O. 1949, The Organization of Behaviour: a neuropsychological theory
  (New York: Wiley and Sons)

\bibitem[{{Hollenbach} \& {McKee}(1989)}]{HollenbachMcKee89}
{Hollenbach}, D. \& {McKee}, C.~F. 1989, \apj, 342, 306

\bibitem[{{Hopfield}(1982)}]{Hopfield82}
{Hopfield}, J.~J. 1982, Proceedings of the National Academy of Science, 79,
  2554

\bibitem[{{Hopfield}(1984)}]{Hopfield84}
---. 1984, Proceedings of the National Academy of Science, 81, 3088

\bibitem[{{Lipovka} {et~al.}(2005){Lipovka}, {N{\'u}{\~n}ez-L{\'o}pez}, \&
  {Avila-Reese}}]{Lipovka05}
{Lipovka}, A., {N{\'u}{\~n}ez-L{\'o}pez}, R., \& {Avila-Reese}, V. 2005,
  \mnras, 361, 850

\bibitem[{{Maio} {et~al.}(2007){Maio}, {Dolag}, {Ciardi}, \&
  {Tornatore}}]{Maio07}
{Maio}, U., {Dolag}, K., {Ciardi}, B., \& {Tornatore}, L. 2007, \mnras, 379,
  963

\bibitem[{{McClelland} {et~al.}(1986){McClelland}, {Rumelhart}, \& the PDP
  Research~Group}]{McClelland86}
{McClelland}, J.~L., {Rumelhart}, D.~E., \& the PDP Research~Group. 1986,
  Parallel Distributed Processing: Explorations in the Microstructure of
  Cognition: Psycological and Bilogical Models, Vol. vol. II (Cambridge, MA:
  MIT Press-Bradford Books)

\bibitem[{{Merlin} {et~al.}(2010){Merlin}, {Buonomo}, {Grassi}, {Piovan}, \&
  {Chiosi}}]{Merlin2010}
{Merlin}, E., {Buonomo}, U., {Grassi}, T., {Piovan}, L., \& {Chiosi}, C. 2010,
  \aap, 513, A36+

\bibitem[{{Merlin} \& {Chiosi}(2007)}]{Merlin2007}
{Merlin}, E. \& {Chiosi}, C. 2007, \aap, 473, 733

\bibitem[{{Prieto} {et~al.}(2008){Prieto}, {Infante}, \&
  {Jimenez}}]{Prieto2008}
{Prieto}, J.~P., {Infante}, L., \& {Jimenez}, R. 2008, ArXiv e-prints

\bibitem[{{Rosenblatt}(1962)}]{Rosenblatt62}
{Rosenblatt}, F. 1962, Principles of neurodynamics; perceptrons and the theory
  of brain mechanisms (Washington: Spartan Books)

\bibitem[{{Rumelhart} {et~al.}(1986){Rumelhart}, {Hinton}, \&
  {Williams}}]{Rumelhart86}
{Rumelhart}, D., {Hinton}, G., \& {Williams}, R. 1986, \nat, 323, 533

\bibitem[{{Sutherland} \& {Dopita}(1993)}]{SutherlandDopita93}
{Sutherland}, R.~S. \& {Dopita}, M.~A. 1993, \apjs, 88, 253

\bibitem[{{Weingartner} \& {Draine}(2001)}]{WeinDraine01}
{Weingartner}, J.~C. \& {Draine}, B.~T. 2001, \apjs, 134, 263

\end{thebibliography}

\end{document}